\documentclass[english]{IEEEtran}
\usepackage[T1]{fontenc}
\usepackage[utf8]{inputenc}
\usepackage{color}
\usepackage{float}
\usepackage{booktabs}
\usepackage{amsmath}
\usepackage{amssymb}

\makeatletter

\providecommand{\tabularnewline}{\\}
\floatstyle{ruled}
\newfloat{algorithm}{tbp}{loa}
\providecommand{\algorithmname}{Algorithm}
\floatname{algorithm}{\protect\algorithmname}

\usepackage[english]{babel}
\usepackage{amsmath}
\usepackage{amsthm}
\usepackage{amsfonts}
\usepackage{amssymb}
\usepackage{graphicx}
\usepackage{tikz}
\usepackage{fixmath}
\usetikzlibrary{bayesnet}
\usepackage{algorithm}
\usepackage{algpseudocode}
\usepackage{pgfplots}

\@ifundefined{showcaptionsetup}{}{%
 \PassOptionsToPackage{caption=false}{subfig}}
\usepackage{subfig}
\makeatother

\usepackage{babel}
\begin{document}

\title{Multiple Instance Learning for Malware Classification}

\author{Jan Stiborek, Tomáš Pevný, Martin Rehák}
\maketitle
\begin{abstract}
This work addresses classification of unknown binaries executed in
sandbox by modeling their interaction with system resources (files,
mutexes, registry keys and communication with servers over the network)
and error messages provided by the operating system, using vocabulary-based
method from the multiple instance learning paradigm. It introduces
similarities suitable for individual resource types that combined
with an approximative clustering method efficiently group the system
resources and define features directly from data. This approach effectively
removes randomization often employed by malware authors and projects
samples into low-dimensional feature space suitable for common classifiers.
An extensive comparison to the state of the art on a large corpus
of binaries demonstrates that the proposed solution achieves superior
results using only a fraction of training samples. Moreover, it makes
use of a source of information different than most of the prior art,
which increases the diversity of tools detecting the malware, hence
making detection evasion more difficult.
\end{abstract}


\section{Motivation}

Since malware is presently one of the most serious threats to computer
security with the number of new samples reaching 140 million in 2015~\cite{avTestMalwareStatistics2016},
battles against it are fought on many fronts. Signature matching remains
the core defense technology, but due to evasion techniques such as
polymorphism, obfuscation, and encryption, keeping good recall is
difficult for static analysis and methods based purely on string matching.
A popular approach to tackle these problems is to execute a binary
in a controlled environment (sandbox)~\cite{Oktavianto2013}, monitor
its behavior, and based on this behavior classify the sample into
benign or malware class (or as a particular malware family). The assumption
of these dynamic analysis methods is that behavior should be more
difficult to randomize and therefore it should constitute a more robust
signal. 

Most approaches to dynamic analysis rely on system calls~\cite{Naval2015,Ahmadi2016,Wuchner2014},
as they are the only means how the binary can interact with the operating
system and other resources. This popularity has however already triggered
many evasion techniques, such as shadow attacks~\cite{Ma2012}, system-call
injection attacks~\cite{Kc2003}, or sandbox detection~\cite{Garcia2016}. 

A perpendicular approach to modeling system calls is to model resources
the binary has interacted with together with the type of the resources.
The rationale is that if malware wants to provide revenue to its owner,
it has to perform\emph{ actions, }such as downloading advertisements
in the case of adware, encrypting hard drive in the case of ransomware,
exfiltrating sensitive data in the case of credential stealers, etc.
This work assumes that execution of these actions involves interactions
with resources visible at the operating system level, and this interaction
can be viewed as a signal which is hard to hide and which can be indicative
of malware families. 

Modeling interactions with system resources has been already exploited
by the prior art. Mohaisen et al.~\cite{Mohaisen2015} extracts a
manually predefined set of features such as number of files created
in specific folders, number of HTTP requests, etc., and use it in
supervised classification. However, we believe that the rapidly changing
threat landscape makes it difficult to manually design features that
are indicative while also being stable over time. An alternative paradigm
is to avoid manual design and to use a bag-of-words model (BoW model),
where every interaction with a particular resource identified by its
name is considered as a unique feature~\cite{Rieck2008}. The price
paid for circumventing manual feature design using BoW is an explosion
of the problem dimension, which can easily reach millions.

This work circumvents the problem of manually designing features while
at the same time avoiding the problem dimension explosion. The approach
is to first cluster resource names with similarity functions tailored
for each resource type (file names, mutexes, registry names, and domain
names), and then use this clustering to represent a sample (a binary
executed in the sandbox) in a lower-dimensional space. This enables
us to use \emph{a random forest classifier} (or any other classifier
of choice) to separate malware from legitimate samples. The clustering
also effectively removes randomization used to evade detection.

The proposed approach is extensively evaluated on a large number of
samples (more than $200\,000$) and compared to relevant prior art.
Experimental results show that the proposed approach indeed improves
the accuracy of detecting malware binaries. 

The contributions of this paper are manifold including a novel approach
to representing malware using raw data, definition of a similarity
measure reflecting directory structure, optimization of similarity
function over binaries, improvements to Louvain clustering in order
to scale to large scale datasets, and finally evaluation and comparison
to state-of-the-art approaches on real-world malware using large-scale
dataset.

\section{Classification of sandboxed samples\label{sec:classification}}

To capture the malware behavior, this work assumes that execution
of malware's actions involves interactions with resources visible
at the operating system level. Examples of such interactions include\emph{
operations with files} during encryption of a victim's hard drive,\emph{
network communication} during data exfiltration or displaying advertisements,
\emph{operation with mutexes} used to ensure a single instance of
malware is running, or manipulation with \emph{registry keys} to ensure
persistency after reboot. An additional source of information are
error messages of the operating system itself. Such information is
provided by the sandboxing environment as the following warnings:\emph{
dll not found} indicating missing dynamic library, \emph{incorrect
executable checksum} indicating corrupted binary, and \emph{sample
did not execute} indicating the fact that the binary was not executed
at all due to various reasons (corrupted binary, sandbox was not able
to copy the binary into VM, etc). 

To model the interactions of a malware binary with resources, this
work views each binary executed in a sandbox as a set of pairs of
names and types of resources the binary interacted with. This view
frames the problem as a \emph{multiple instance learning} (MIL) problem
where each sample (binary) consists of a set of \emph{instances} of
different size. In our scenario an instance represents the pair of
name and type of a resource the binary interacted with during sandboxing. 

\begin{algorithm}
\begin{algorithmic}[1]
\Function{train}{$S,y$}\Comment{Training samples and labels}
\State{$I \leftarrow \textit{extractInstances}(S)$}
\State{$C \leftarrow \textit{cluster}(I)$}\Comment{\parbox[t][3.6em]{.45\linewidth}{Clustering of instances (separately for individual types)}}
\State{$X \leftarrow \textit{project}(S, C)$}\Comment{\parbox[t][2.4em]{.45\linewidth}{Projection of samples into binary vector (Alg.~\ref{alg:project})}}
\State{$\mathcal{M} \leftarrow \textit{trainClassifier}(X, y)$}
\State{\Return{$\mathcal{M},C$}}\Comment{\parbox[t][2.4em]{.45\linewidth}{Returns cluster centers $C$ and trained classifier $\mathcal{M}$}}
\EndFunction

\Function{predict}{$S',C,\mathcal{M}$}\Comment{\parbox[t][2.4em]{.45\linewidth}{Testing samples $S'$, clusters $C$ and classifier $\mathcal{M}$}}
\State{$X' \leftarrow \textit{project}(S',C)$}\Comment{\parbox[t][2.4em]{.45\linewidth}{Projection of samples into binary vector (Alg.~\ref{alg:project})}}
\State{$\hat{y} \leftarrow \textit{predict}(\mathcal{M},X')$} \Comment{\parbox[t]{.45\linewidth}{Classification of testing samples}}
\State{\Return{$\hat{y}$}}
\EndFunction
\end{algorithmic}

\caption{High-level overview of training (function TRAIN) and classification
(function PREDICT) of malware samples.}

\label{alg:overview}
\end{algorithm}
\begin{algorithm}
\begin{algorithmic}[1]
\Function{project}{$S,C$}\Comment{Samples $S$ and clusters $C$.}
\State{$X\leftarrow \emptyset$}
\ForAll{$s\in S$}
	\State{$I \leftarrow \textit{extractInstances}(s)$}
	\State{$x\leftarrow \vec{0}$}
	\ForAll{$i \in I$}
		\State{$c^* \leftarrow \textit{nnSearch}(i,C)$}\Comment{\parbox[t][2.4em]{.35\linewidth}{Finds closest center $c^*$ to instance $i$.}}
		\State{$x[c^*] \leftarrow 1$}
	\EndFor
	\State{$X \leftarrow X \cup \{x\}$}
\EndFor
\State{\Return{$X$}}
\EndFunction
\end{algorithmic}

\caption{Projection of samples $S$ into binary vector using cluster centers
$C$.}

\label{alg:project}
\end{algorithm}
Variable sizes of samples and lack of order over their instances pose
a challenge to traditional machine learning methods that expect samples
to have fixed size. A recent review of MIL algorithms~\cite{Amores2013}
lists various approaches to overcome this variability in sample sizes.
One of the most popular (also adopted in this work) is vocabulary-based
method outlined in Algorithm~\ref{alg:overview}. It employs clustering
of instances to describe the sample by a fixed-dimensional vector
with length equal to the size of vocabulary, i.e. a set of clusters,
so that an ordinary machine learning method can be applied.

To convert a sample into a fixed-dimensional vector, all instances
$I$ from all training samples $S$ are extracted and clustered by
a suitable method per given resource type\textendash files, mutexes,
registry keys, network communication. Note that warnings generated
by the sandboxing environment are used directly, i.e. every warning
is considered as a separated cluster. The resulting clusters represent
the vocabulary. Next, for every instance $i$ the closest cluster
prototype $c^{*}$ (a small random subset of the cluster of instances)
of corresponding type is located. Finally, the binary representation
is then used such that element of the vector equals to 1 iff there
was an instance close to the particular cluster prototype. Once all
samples are encoded as fixed-dimensional vectors, one can use a machine
learning algorithm of choice to implement the classifier. This work
uses the \emph{random forest classifier}~\cite{Breiman2001} due
to its versatility, accuracy, and scalability, which make it a popular
choice for many different machine learning tasks including malware
classification \cite{Hansen2016}.

Since the clustering is an essential component of the above algorithm,
the definition of similarity over instances (resource names) greatly
influences the accuracy of the system, and therefore it should reflect
properties of the application domain. The rest of this section defines
a specific similarity metric for each type of resources the malware
interact with, namely on files, mutexes, network hostnames, and registry
keys, and also justifies our choice of the clustering method. 

\subsection{Similarity between file paths\label{subsec:File-path-preprocessing}}

\begin{table}
\begin{centering}
\subfloat[\label{tab:artifactsSamples} raw filenames]{\begin{centering}
\begin{tabular}{ll}
\toprule 
Binary 1 & Binary 2\tabularnewline
\midrule
\textbackslash{}Temp\textbackslash{}4ffdd6ab-8020\textbackslash{}config.dmc & \textbackslash{}Temp\textbackslash{}ed8a9718-c7a0\textbackslash{}config.dmc\tabularnewline
\textbackslash{}Temp\textbackslash{}4ffdd6ab-8020\textbackslash{}bin.dmc & \textbackslash{}Temp\textbackslash{}ed8a9718-c7a0\textbackslash{}bin.dmc\tabularnewline
\textbackslash{}Windows\textbackslash{}System32\textbackslash{}ftp.exe & \textbackslash{}Windows\textbackslash{}System32\textbackslash{}netsh.exe\tabularnewline
\bottomrule
\end{tabular}
\par\end{centering}
}
\par\end{centering}
\begin{centering}
\subfloat[\label{tab:artifactSamples_translated} artifact clusters]{\begin{centering}
\begin{tabular}{ll}
\toprule 
Binary 1 & Binary 2\tabularnewline
\midrule
Artifact cluster 1 & Artifact cluster 1\tabularnewline
Artifact cluster 1 & Artifact cluster 1\tabularnewline
Artifact cluster 2 & Artifact cluster 2\tabularnewline
\bottomrule
\end{tabular}
\par\end{centering}
}
\par\end{centering}
\caption{Example of clustering files of two binaries from the same family executed
in the sandbox.}
\end{table}
Although viewing file paths as strings would allow to use vast prior
art such as Levenshtein distance~\cite{Levenshtein1966}, Hamming
distance, Jaro-Winkler distance~\cite{Navarro2001}, or string kernels
introduced in \cite{Lodhi2002}, the file systems were designed as
tree structures with names of some folders (fragments of the path)
being imposed by the operating system and the distance should reflect
that. For example two files with paths \texttt{/Documents and Settings/Admin/Start
Menu/Programs/Startup/tii9fwliiv.lnk} and\texttt{ /Documents and Settings/Admin/Start
Menu/Programs/Accessories/Notepad.lnk} share large parts of their
paths and common string similarities will return high similarity score,
but they serve very different purposes, since the first file is a
link to an application executed after the start of the operating system
(OS), while the second is a regular link in the Start menu in Windows
OS. Another aspect that prohibits the use of common string similarities
is their computational complexity (typically $O(n^{2})$ where $n$
is the length of the string). The complexity combined with the number
of resources to be clustered (in order of millions) leads to unfeasible
time requirements. This motivates the design of a similarity that
is fast and takes into the account the tree structure of the file
system, special folders, and differences between folders and filenames.

The proposed similarity $s(x,x')$ of two file paths $x$ and $x^{\prime}$
is defined as

\begin{equation}
s(x,x')=\exp\left(-w^{T}f(x,x')\right),\label{eq:fileSimilarity}
\end{equation}
where $w$ is a vector of weights and $f(x,x^{\mathrm{\prime}})$
is a function extracting a feature vector from file paths $x$ and
$x'.$ Both the weight vector $w$ and function $f$ play an essential
role and are both discussed in detail below.

The function $f$ in~(\ref{eq:fileSimilarity}) captures differences
between the two paths $x$ and $x'$ by a fixed-dimensional vector.
It first splits both paths $x$ and $x^{\prime}$ into fragments $x_{i}$
and $x'_{i}$ using OS specific path separator\footnote{Unixes and MacOS uses '/' as a path separator, Windows uses '\textbackslash{}'.},
in the cases of MacOS and Windows changes all characters to lowercase,
and assigns all fragments into one of the following four categories:
\begin{enumerate}
\item \emph{known folder }\textendash{} fragment $x_{i}$ is a well known
folder in the list of folders imposed by the operation system (e.g.
\texttt{Windows}, \texttt{Program Files}, \texttt{System32}, etc.),
\item \emph{general folder }\textendash{} fragment $x_{i}$ is a not-well-known
folder (e.g. unknown folders in \texttt{Program Files}, randomly generated
folders in Internet Explorer cache folder, etc.),
\item \emph{file }\textendash{} fragment $x_{i}$ is file,
\item \emph{empty }\textendash{} artificial fragment used for padding the
paths in cases when paths $x$ and $x^{\prime}$ have different depths.
\end{enumerate}
When all fragments are assigned to one of the above classes, their
dissimilarity is captured by the function $f$ as 
\begin{align*}
f(x,x^{\prime}) & =(f_{KK},f_{KG},f_{KF},f_{KE},f_{GG},f_{GF},f_{GE},f_{FF},f_{FE})
\end{align*}
 where
\begin{itemize}
\item $f_{KK}$ is the number of fragments on the same level that were both
classified as \emph{known folder }and were not equal,
\item $f_{GG}$ is the sum of Levenshtein distances between all fragments
on the same level that were classified as \emph{general folder},
\item $f_{FF}$ is the sum of Levenshtein distances of all fragments on
the same level that were classified as \emph{file},
\item $f_{KG}$, $f_{KF}$, $f_{KE}$, $f_{GF}$, $f_{GE}$, $f_{FE}$ are
the sums of all fragments of the same level and were classified as
\emph{known} and \emph{general folder, known folder }and \emph{file,
known folder }and \emph{empty, general folder }and \emph{file, general
folder }and \emph{empty, }and \emph{file }and \emph{empty} respectively.
\end{itemize}
\begin{table*}
\begin{centering}
\begin{tabular}{lllllll}
\toprule 
 & Fragment 1 & Fragment 2 & Fragment 3 & Fragment 4 & Fragment 5 & Fragment 6\tabularnewline
\midrule
$x$ & \texttt{Documents and Settings} ($K$) & \texttt{Admin} ($G$) & \texttt{Start Menu} ($K$) & \texttt{Programs} ($K$) & \texttt{Startup} ($K$) & \texttt{tii9fwliiv.lnk} ($F$)\tabularnewline
$x'$ & \texttt{Documents and Settings} ($K$) & \texttt{Admin} ($G$) & \texttt{Start Menu} ($K$) & \texttt{Programs} ($K$) & \texttt{Accessories} ($G$) & \texttt{Notepad.lnk} ($F$)\tabularnewline
\bottomrule
\end{tabular}
\par\end{centering}
\caption{Example of two paths $x$ and $x'$ separated into individual fragments
with labels ($K$ \textendash{} known folder, $G$ \textendash{} general
folder and $F$ \textendash{} file).}
\label{tab:fragmentClassifExample}
\end{table*}
To illustrate the calculation of $f(x,x')$, let's consider the same
two paths used above. At first, function $f$ splits both paths into
fragments and assign them into one of four categories (see Table~\ref{tab:fragmentClassifExample}).
Assigning fragment to classes requires a list of known folders\footnote{Full list of known folders is available online: https://github.com/SfinxCZ/Malware-analysis-using-multiple-instance-learning},
which for the purpose of this example we assume to contain \texttt{Documents
and Settings}, \texttt{Start Menu}, \texttt{Programs} and\texttt{
Startup}, which are present in all windows installations. All corresponding
folders from those two paths are therefore assigned to known folder
class, while \texttt{Admin} and \texttt{Accessories} are labeled as
general folders.\footnote{The first three known folders are embedded in the functionality of
the Windows OS. The \texttt{Startup} folder has a specific meaning
altering the behavior of the operation system since all programs listed
in this folder are executed after the boot of the OS. On the other
hand \texttt{Accessories} can be easily changed without major consequences.} Individual elements of the vector $f(x,x')$ are calculated using
the above rules as follows: the first rule applies to three fragments
1, 3, and 4 belonging to known folder class, but as they are all equal
$f_{KK}=0;$ the second rule returns $0$ based on analogous reasoning
but for general folders; the third rule returns $f_{FF}=0.7143,$
which is the Levenshtein distance between \texttt{tii9fwliiv.lnk}
and \texttt{Notepad.lnk}; the only mismatch is on fragment 5\textendash known
folder and general folder yielding $f_{KG}=1;$ and finally all remaining
elements of feature vector are $0$. The output of $f(x,x')$ is captured
by the feature vector

\[
f(x,x')=(0,0,0.7143,1,0,0,0,0,0).
\]

The weight vector $w$ in~(\ref{eq:fileSimilarity}) captures the
contribution of individual elements of the feature vector $f(x,x')$.
Imposing condition $w\ge0$, in combination with construction of function
$f$, bounds the value of the similarity function~(\ref{eq:fileSimilarity})
$s(x,x')\in\left[0,1\right]$ such that the similarity functions returns
$1$ (or values close to $1$) if $x$ and $x^{\prime}$ belong to
the same class (files in \texttt{/temp/}\emph{ }directory, cache of
the Internet Explorer, files in system directory, etc.) and values
approaching $0$ if they belong to different classes. Since the similarity
function~(\ref{eq:fileSimilarity}) was inspired by the popular Gaussian
kernel, the parameter vector $w$ was optimized using the Centered
Kernel Target Alignment~\cite{Cortes2012} (CKTA), which is a method
to optimize kernel parameters. CKTA assumes training data $\lbrace(x_{i},y_{i}),\rbrace_{i=1}^{m}$
where $x_{i}$ is a file path and $y_{i}$ is the class of the path
$x_{i}$, and defines \emph{centered kernel matrix} as 

\begin{align}
\left[\mathbf{S}_{c}^{w}\right]_{ij}=\mathbf{S}_{ij}^{w}-\frac{1}{m}\sum_{i=1}^{m}\mathbf{S}_{ij}^{w}-\frac{1}{m}\sum_{j=1}^{m}\mathbf{S}_{ij}^{w}+\frac{1}{m^{2}}\sum_{i,j=1}^{m}\mathbf{S}_{ij}^{w},\label{eq:centeredKernelMat}
\end{align}
where $\text{S}_{ij}^{w}=s_{w}(x_{i},x_{j})$ is the kernel matrix
corresponding to the similarity function~(\ref{eq:fileSimilarity})
parametrized by the weight vector $w.$ CKTA maximizes correlation
between labels and a similarity matrix by solving the following optimization
problem

\begin{equation}
w^{*}=\arg\max_{w\ge0}\frac{\left\langle \mathbf{S}_{c}^{w},\mathbf{Y}{}_{c}\right\rangle _{F}}{\left\Vert \mathbf{S}_{c}^{w}\right\Vert _{F}\cdot\left\Vert \mathbf{Y}_{c}\right\Vert _{F}},\label{eq:CKTAoptimization}
\end{equation}
where $\mathbf{Y}$ is target label kernel with $\left[\mathbf{Y}\right]_{ij}$
equals to 1 when $i^{\text{th}}$ and $j^{\text{th}}$ paths from
training data belongs to the same class and $-1$ otherwise, $\left\langle \cdot,\cdot\right\rangle _{F}$
is Frobenius product and $\left\Vert \cdot\right\Vert _{F}$ is Frobenius
norm (see Appendix~\ref{subsec:Frobenius-product} for more details).
In below experiments~(\ref{eq:CKTAoptimization}) is solved by \emph{stochastic
gradient descent (SGD) }algorithm~\cite{Bishop2011}. Note that although
the path similarity $s(x_{i},x_{j})$ is not a valid kernel because
it is not positive definite, the use of centered kernel alignment
is still possible as the only limitation is that the global optimum
might not be found. 

To finish the example, the similarity function (\ref{eq:fileSimilarity})
with weight vector $w=(2,10^{-5},1,2.3,1.6,1,0.36,0.7,0.9)$ returns
the value $s(x,x')=0.049,$ which correctly indicates that the two
paths are different.

\subsection{Similarity of network traffic}

To define the similarity between network resources one has to overcome
the randomization often employed by malware authors that render trivial
similarity based on names of network resources (domains, IPs) ineffective.
To escape blacklisting command and control (C\&C) channels of malware,
its authors use various techniques to hide and obscure C\&C operation.
Popular approaches include randomization of domain names by generating
them randomly (DGA), quickly changing hosting servers and / or domain
names by  fast flux, or using large hosting providers like Amazon
Web Services to hide among legitimate servers, etc. These techniques
are relatively cheap (e.g. registering a new \emph{.com} domain costs
\textasciitilde{}3USD per 1 year) and they allow for variation in
domain names without updating disseminated malware binaries. In contrast,
switching from one C\&C paradigm to another requires such an update
and therefore occurs relatively infrequently. These two properties
contribute to each malware family using specific patterns of domain
names, paths, and parts of URLs. Exploiting these patterns allows
to group domain names into clusters. In this work the similarity in
network traffic is defined only for HTTP/HTTPS protocol, because it
is presently the default choice for malware authors as it is rarely
filtered. The extension to other network traffic is possible~\cite{Kohout2015}.

The similarity in URL patterns used in this work has been adopted
from~\cite{Jusko2016}, which has proposed to cluster domain names
so that each cluster contains domains of one type / for one family
of malware. The calculation of similarity starts by grouping all HTTP/HTTPS
requests using the domain names. Then the model of each domain name
is built from path and query strings, transferred bytes, duration
of requests and inter-arrival times (time spans between requests to
the same domain) of individual requests to it. Finally, these models
are used to calculate the similarity function between two domain names
in the clustering. Since the calculation of the similarity is out
of scope, we refer to an original publication~\cite{Jusko2016} for
details.

\subsection{Similarity between mutex names}

\emph{Mutex} (\emph{Mutual exclusive object}) is a service provided
by most modern operating systems to synchronize multi-threaded and
multi-processes applications. This mechanism is popular among malware
authors to prevent multiple infections of the same machine, because
running two instances of the same malware can cause conflicts limiting
the potential revenue. Mutexes are identified by their name, which
can be an arbitrary string. The naming scheme is challenging for malware
authors, because the names cannot be static, which would make them
good indicators of compromise of a particular malware, but they cannot
be completely random either, because two independent binaries of the
same family would not be able to check the presence of each other.
Therefore malware authors resorted to pseudo-deterministic algorithms
or patterns for generating mutex names. For some malware families
these patterns are already well known, for example Sality~\cite{SymantecSecurityResponse2011}
uses mutex names of the form \texttt{\textquotedbl{}<process name>.exeM\_<process
ID>\_\textquotedbl{}}\texttt{\emph{-}}\texttt{explorer.exeM\_1423\_}.

Since operating systems do not impose any restrictions on the names
of mutexes, they can be arbitrary strings. Therefore standard string
similarities such as Levenshtein distance, Hamming distance, Jaro-Winkler
distance, etc. can be used. In experiments presented in Sections~\ref{sec:Experiments}
Levenshtein was used, as it gives overall good results. 

\subsection{Similarity between registry names}

In Microsoft Windows operating system, the primary target of the majority
of malware, registry serves as a place where programs can store various
configuration data. It is a replacement of configuration files with
several improvements such as strongly typed values, faster parsing,
ability to store binary data, etc. The registry is a key-value store,
where key names have the structure of a file system. The root keys
are \texttt{HKEY\_LOCAL\_MACHINE}, \texttt{HKEY\_CURRENT\_USER},\emph{
}\texttt{HKEY\_CURRENT\_CONFIG}, \texttt{HKEY\_CLASSES\_ROOT}, \texttt{HKEY\_USERS}\emph{
}and\emph{ }\texttt{HKEY\_PERFORMANCE\_DATA}; some root keys also
always have sub-keys with specific names (\texttt{Software}, \texttt{Microsoft},
\texttt{Windows}, etc.)\emph{. }Due to similarity with a file system,
the similarity distance is the same as the one defined in Subsection~\ref{subsec:File-path-preprocessing},
but with a different set of names of known folders and a weight vector
optimized on registry data rather than on files.

\subsection{Clustering of resource names\label{subsec:clusteringDescription}}

The above similarities are not true distances, which limits the choice
of applicable clustering methods to those that do not require proper
distance metric between points. The Louvain method~\cite{Blondel2008}
is a popular choice and it is used in experiments below, because it
also automatically determines the number of clusters and thus removes
the need to set it manually. The use of the Louvain method is the
authors' preference, but other clustering methods can be used as well;
the reader is referred to~\cite{Fortunato2010} for an overview of
methods requiring only similarity. 

\begin{algorithm}
\begin{algorithmic}[1]
\Function{approxCluster}{$I;k,m,\epsilon$}
\State{$C=\emptyset$}
\While{$I \neq \emptyset$} 
  \State{$I'\leftarrow$ Random subset of size $k$ from $I$}
  \State{$C' \leftarrow \textit{cluster}(I',m)$}\Comment{\parbox[t][3.6em]{.35\linewidth}{Cluster instances $I'$ and create cluster prot. of size $m$.}}
  \ForAll{$i \in I\setminus I'$}
    \State{$c^* \leftarrow \textit{nnSearch}(i,C')$}\Comment{\parbox[t][2.4em]{.35\linewidth}{Find cluster prot. $c^*$ closest to instance $i$.}}
	\If {$s(i,c^*)>\epsilon$}
		\State{$c^* \leftarrow c^* \cup \{i\}$}
	\EndIf
  \EndFor
  \State{$C \leftarrow C\cup C'$}
\EndWhile
\State{\Return{$C$}}
\EndFunction
\end{algorithmic}

\caption{Approximative clustering algorithm for instances $I$ (resource names).}

\label{alg:clustering}
\end{algorithm}
The use of the Louvain method is not straightforward in the scenario
of this paper because it requires a full adjacency matrix in advance.
This results in a lower bound to computational complexity being $O(n^{2})$
in the number of resources, which is clearly prohibitive as the number
of unique resource names to cluster can easily reach the order of
millions. To decrease the number of calculated similarities, an approach
inspired by~\cite{Zhang1997,Zimek2013,Jang2010} is adopted where
the Louvain clustering is used iteratively as summarized in Algorithm~\ref{alg:clustering}.
Given a set of instances $I$ of a particular type, in every iteration
the algorithm selects a random subset $I'\subset I$ of the data of
size $k$ small enough for the Louvain method to be computationally
feasible. The results of the Louvaine clustering are then transformed
to cluster prototypes\textemdash random subsets of clusters with size
limited to $m$. Remaining data $I\backslash I'$ are then traversed
and all samples with similarity larger than $\epsilon$ to some cluster
prototype $c^{*}\in C'$ are added to $c^{*}$ and removed from $I.$
Finally, $C'$ is merged with the clustering $C$ obtained in the
previous iteration, and if $I$ is not empty, the process is repeated.

Clearly the algorithm is an approximation of a clustering with complete
data and its performance depends on the choice of parameters $k$
and $\epsilon$. Experiments indicate that if parameter $k$ is large
enough ($k=10^{5}$) and parameter $\epsilon$ is set reasonably (in
the experimental evaluation we use $\epsilon=0.4$, see Section~\ref{subsec:ParamOpt}
for details), the results are comparable with clustering methods applied
to the complete data. The computational complexity of this sequential
approximation is $O\left(l\cdot\left(k\cdot\left(k-1\right)/2+c_{l}\cdot m\cdot\left(n_{l}-k\right)\right)\right)$
where $l$ is the number of iterations of algorithm (typically $l\leq10$),
$n_{l}$ is the number of non-clustered samples in $l$-th iteration,
$k$ is the number of randomly selected samples, $c_{l}$ is the number
of cluster prototypes produced by the clustering algorithm in $l$-th
iteration and $m$ is the maximal size of a cluster prototype (typically
$m=10$). Since the parameter $k$ is fixed and $k\ll n$, we can
see that the number of evaluations of the similarity function is linear
in the number of samples, which clearly outperforms the quadratic
complexity required by the vanilla Louvain method.

\section{Evaluation\label{sec:Experiments}}

In this section the proposed approach is compared to the approach
proposed by Rieck, et al.~\cite{Rieck2008} (further referred to
as \emph{Rieck}) and the approach proposed by Mohaisen, et al. \cite{Mohaisen2015}
(further referred to as \emph{AMAL}). Rieck has been selected as a
representative of the prior art that encodes malware behavior into
a high-dimensional feature space using bag-of-words model built directly
from data; it uses kernelized SVM to classify binaries. The second
approach, AMAL, encodes malware behavior using a relatively low number
of hand-made features; to classify unknown binaries AMAL trains multiple
classifiers (SVM, decision trees, k-nearest neighbor, etc.) and selects
the optimal classifier for given data using cross-validation. 

\subsection{Data set description}

\begin{table}
\centering{}%
\begin{tabular}{lrclr}
\toprule 
Malware family & \#samples &  & Malware family & \#samples\tabularnewline
\midrule
nemucod & $13\,781$ &  & amonetize & $1172$\tabularnewline
cerber & $12\,829$ &  & nanocore & $1032$\tabularnewline
bladabindi & $10\,945$ &  & loadmoney & $964$\tabularnewline
locky & $9894$ &  & yakes & $892$\tabularnewline
gamarue & $7694$ &  & bifrose & $804$\tabularnewline
darkkomet & $4664$ &  & autoit & $781$\tabularnewline
hupigon & $3555$ &  & kolabc & $707$\tabularnewline
upatre & $3269$ &  & waldek & $686$\tabularnewline
tinba & $3104$ &  & pdfka & $649$\tabularnewline
scar & $2961$ &  & shipup & $625$\tabularnewline
swrort & $2868$ &  & rebhip & $613$\tabularnewline
zbot & $2426$ &  & razy & $599$\tabularnewline
virlock & $1797$ &  & agentb & $579$\tabularnewline
fareit & $1763$ &  & poison & $551$\tabularnewline
farfli & $1749$ &  & xtrat & $511$\tabularnewline
zegost & $1719$ &  & onlinegames & $502$\tabularnewline
virut & $1556$ &  & ramnit & $493$\tabularnewline
adwind & $1537$ &  & magania & $463$\tabularnewline
zusy & $1505$ &  & atraps & $461$\tabularnewline
ircbot & $1447$ &  & softpulse & $460$\tabularnewline
zerber & $1329$ &  & banload & $387$\tabularnewline
palevo & $1270$ &  & ruskill & $374$\tabularnewline
vobfus & $1244$ &  & downloadassistant & $373$\tabularnewline
delf & $1228$ &  & binder & $350$\tabularnewline
donoff & $1211$ &  & \emph{remaining MW families} & $31\,856$\tabularnewline
\midrule 
Total malicious & \multicolumn{4}{r}{$144\,229$}\tabularnewline
\midrule 
Total legitimate & \multicolumn{4}{r}{$87\,026$}\tabularnewline
\bottomrule
\end{tabular}\caption{\label{tab:MalwareFamilies}Number of samples of malware families
in the data set. The malware families for individual samples were
determined using AVClass tool~\cite{Sebastian2016}.}
\end{table}
\begin{table}
\begin{centering}
\begin{tabular}{ll}
\toprule 
AhnLab, \emph{V3 Internet Security} & G Data, \emph{InternetSecurity}\tabularnewline
Avira, \emph{Antivirus Pro} & Kaspersky Lab, \emph{Internet Security}\tabularnewline
Bitdefender, \emph{Internet Security} & Microworld, \emph{eScan internet security suite}\tabularnewline
ESET, \emph{Internet Security} & Symantec, \emph{Norton Security}\tabularnewline
F-Secure, \emph{Safe} & Trend Micro,\emph{ Internet Security}\tabularnewline
\bottomrule
\end{tabular}
\par\end{centering}
\caption{\label{tab:selAVEngines}Selected AV engines that received full 6
points for performance in AV-Test report from December 2016 \cite{AV-Test2016}.}

\end{table}

The dataset used for experiments contained $250\,527$ files collected
from October 24, 2016 to December 12, 2016 using \emph{AMP ThreatGrid}~\cite{ampThreatGrid}.
All files were also analyzed by \emph{VirusTotal.com} service~\cite{virustotal}
and labeled using its verdicts as follows: a file was labeled as malicious
if at least 4 out of 10 selected AV engines (see Table~\ref{tab:selAVEngines}
for details) detected the file as malicious, and it was labeled as
legitimate if none of the AV engines detected the file. Remaining
files were discarded as unknown and removed from both training and
testing sets in order to limit the effect of misclassifications by
individual AV engines. The final numbers of files were: $144\,229$
malicious, $87\,026$ legitimate, and $19\,272$ discarded as unknown.
The numbers of samples of individual malware families are summarized
in Table~\ref{tab:MalwareFamilies}.

All files were executed in sandbox by AMP ThreatGrid~\cite{ampThreatGrid}
service, using Windows 7 64bit ($71\%$ samples) environment, as it
is the most popular OS at the time of writing\footnote{According to http://www.w3schools.com/browsers/browsers\_os.asp Windows
7 has 34.6\% market share against 1.0\% covered by Windows XP, 11.1\%
covered by Windows 8 and 30.9\% covered by Windows 10.}, and Windows XP ($29\%$ samples) environment, since it is still
widely deployed on embedded machines such as ATMs. Virtual machines
were connected to the Internet without any filtering or restrictions
that could by any mean prevent connections to command \& control servers
or other servers. The work here is not tailored to AMP ThreatGrid,
as the same or similar information about binaries can be obtained
by a number of different sandboxing solutions such as Cuckoo \cite{Oktavianto2013},
Ether \cite{Dinaburg2008}, or CWSandbox \cite{Willems2007}.

In contrast to the majority of prior art, binaries were divided into
training and testing sets according to the dates they were collected
rather than randomly. This approach is more realistic since it does
not overestimate the detection performance as some malware families
may not be known at the time of training, as they might have appeared
later. Thus, all training samples collected prior to November 12,
2016 ($72\,963$ malicious binaries and $48\,152$ legitimate binaries)
were used for training, and remaining samples ($71\,266$ malicious
binaries and $38\,874$ legitimate binaries) were used for testing.

\subsection{Hyper-parameter optimization\label{subsec:ParamOpt}}

All compared methods have several parameters that have to be tuned
to achieve good detection accuracy. While in Rieck and the proposed
method the parameters have to be optimized using grid search (detailed
below), AMAL is designed to perform such optimization during training
in order to select both the optimal classifier (SVM, linear SVM, decision
trees, logistic regression, k-nearest neighbor and perceptron) and
its parameters and thus it does not need to optimize its parameters
in advance. 

Since Rieck uses SVM with L2 regularization and polynomial kernel
there are two parameters that need to be tuned: misclassification
cost $C\in\left\{ 10^{-2},\dots,10^{8}\right\} $ and degree of the
kernel $d\in\left\{ 1,\dots,5\right\} $. The optimal configuration
achieving highest accuracy estimated by five-fold cross-validation
on the training data was $C=10^{4},d=4$.

The random forest classifier described in Section~\ref{sec:classification}
contains several parameters such as the number of trees $K\in\left\{ 10,20,50,100,200\right\} $,
maximal depth $d_{m}\in\left\{ 5,10,30,50,\infty\right\} $, minimal
number of samples in node to perform split $s_{n}\in\left\{ 2,4,6,10,20\right\} $,
and criterion $c\in\left\{ \text{gini},\text{entropy}\right\} $.
All remaining parameters (maximal number of features, minimal number
of samples in leaf, maximal number of leafs, class weights, minimum
weighted fraction of the total sum of weights in leaf, minimal impurity
for split) were set to their default values as defined in the Scikit-learn
library \cite{Pedregosa2011} since according to our experiments they
have little influence on detection performance. The optimal configuration
of parameters with respect to accuracy estimated by five-fold cross-validation
on training data was $K=100$, $d_{m}=\infty$, $s_{n}=2$ and $c=\text{gini}$. 

Additional two parameters (size of randomly selected subsets $k\in\left\{ 10^{4},2\cdot10^{4},5\cdot10^{4},10^{5},2\cdot10^{5},5\cdot10^{5},\infty\right\} $
and minimal similarity $\epsilon\in\left\{ 0.1,\dots,0.9\right\} $)
affect the clustering of the resource names described in Section~\ref{subsec:clusteringDescription}.
The minimal similarity was optimized on a manually labeled set of
file paths and registry keys that were clustered with different values
of $\epsilon$. The resulting clusters were evaluated with respect
to the \emph{adjusted rand index \cite{Rand1971},} a well known score
for evaluation of clustering algorithms, and the optimal value of
$\epsilon=0.4$ was selected. To find the optimal size of randomly
selected subsets $k$ the accuracy of the whole proposed method with
different settings of parameter $k$ was estimated using five fold
cross validation on randomly selected subset of training data\footnote{The subset was limited to \textasciitilde{}30 000 samples in order
to limit the number of resources so that complete clustering could
be performed.}. Since the differences between various settings were negligible,
the value of the parameter $k=10^{5}$ was selected as a reasonable
balance. Low value of parameter $k$ increases the number of iterations
$l$ performed by the clustering algorithm, since too many samples
are rejected to be too dissimilar to available cluster prototypes,
and high value increases the quadratic cost for computation of adjacency
matrix required by Louvain method.

Classification performance was measured with standard evaluation metrics
\cite{Fawcett2006a}: \emph{true positive rate (TPR), false negative
rate (FNR),} \emph{true negative rate (TNR), false positive rate (FPR)}
and \emph{accuracy}. Since the experimental scenario is binary (positive
malware vs. negative benign), the TPR (FNR) is the proportion of correctly
(incorrectly) classified malware samples, TNR (FPR) is the proportion
of correctly (incorrectly) classified legitimate samples and accuracy
is the rate of correctly classified samples regardless their class. 

\subsection{Experimental results}

The comparison and evaluation is divided into two parts. The first
experiment evaluates the detection performance of the proposed method,
Rieck and AMAL trained on the full training set ($121\,115$ samples),
while the second experiment measures degradation of detection performance
when only a limited number of data are available for training ($5\%,$
$10\%$, $20\%$ and $100\%$ of training samples). Note that to evaluate
AMAL on the complete training set, the meta learner was not allowed
to use SVM classifier with RBF kernel due to excessive computational
requirements. Note that AMAL's meta-learner has never selected this
variant of the SVM classifier in smaller experiments performed in
this work, hence removing it most probably does not have any impact
on the results. 

\begin{table}[t]
\begin{centering}
\begin{tabular}{lccccccc}
\toprule 
 & \multicolumn{3}{c}{estimated on testing set} &  & \multicolumn{3}{c}{estimated on training set}\tabularnewline
\cmidrule{2-4} \cmidrule{6-8} 
 & TPR & FPR & ACC &  & TPR & FPR & ACC\tabularnewline
\midrule
This paper & \textbf{0.954} & \textbf{0.067} & \textbf{0.943} &  & 0.973 & 0.061 & 0.956\tabularnewline
Rieck & 0.934 & 0.081 & 0.926 &  & \textbf{0.974} & \textbf{0.014} & \textbf{0.980}\tabularnewline
AMAL & 0.795 & 0.108 & 0.845 &  & 0.845 & 0.047 & 0.899\tabularnewline
\bottomrule
\end{tabular}
\par\end{centering}
\caption{True (TPR) and false (FPR) positive rates of evaluated methods estimated
on the training and testing set.}
\label{tab:results}
\end{table}
\begin{figure*}[t]
\subfloat[False negative rates]{\begin{tikzpicture} 
\begin{axis}[     
xlabel=Size of training set,     
ylabel=FNR,  
ymin=0,
ymax=0.15,
xtick={1,2,3,4},     
xticklabels={$5\%$,$10\%$,$20\%$,$100\%$},     
legend style={at={(0.5,1.1)},anchor=south},      	
legend columns=-1,      	
scaled ticks=false, 	
tick label style={ 	
	/pgf/number format/fixed, 	
	/pgf/number format/.cd,
    fixed zerofill,
    precision=3 
	}      
] 
\addplot[dashed,error bars/.cd, y dir = both, y explicit] table [x=i,y=fnr,y error=fnr_std, col sep=comma] {figs/fnr_rieck.csv}; 
%\addplot[dashdotted,error bars/.cd, y dir = both, y explicit] table [x=i,y=fnr,y error=fnr_std, col sep=comma] {figs/fnr_amal.csv};
\addplot[solid,error bars/.cd, y dir = both, y explicit] table [x=i,y=fnr,y error=fnr_std, col sep=comma] {figs/fnr_rf.csv};
\addlegendentry{Rieck} 
%\addlegendentry{AMAL}
\addlegendentry{Proposed approach}
\end{axis} 
\end{tikzpicture}

\label{fig:fnr_degradation}}\subfloat[False positive rate]{\begin{centering}
\begin{tikzpicture} 
\begin{axis}[     
xlabel=Size of training set,     
ylabel=FPR,
ymin=0,
ymax=0.15,
xtick={1,2,3,4},
xticklabels={$5\%$,$10\%$,$20\%$,$100\%$},     
legend style={at={(0.5,1.1)},anchor=south},      	
legend columns=-1,      	
scaled ticks=false,  	
tick label style={ 	
	/pgf/number format/fixed, 	
	/pgf/number format/.cd,     
		fixed zerofill,     
		precision=3 	
}      
] 
\addplot[dashed,error bars/.cd, y dir = both, y explicit] table [x=i,y=fpr,y error=fpr_std, col sep=comma] {figs/fpr_rieck.csv};
%\addplot[dashdotted,error bars/.cd, y dir = both, y explicit] table [x=i,y=fpr,y error=fpr_std, col sep=comma] {figs/fpr_amal.csv}; 
\addplot[solid,error bars/.cd, y dir = both, y explicit] table [x=i,y=fpr,y error=fpr_std, col sep=comma] {figs/fpr_rf.csv};
\addlegendentry{Rieck}
%\addlegendentry{AMAL}
\addlegendentry{Proposed approach}
\end{axis} 
\end{tikzpicture}
\par\end{centering}
\label{fig:fpr_degradation}}\caption{Comparison of FNR and FPR for Rieck and proposed method trained on
training sets of different sizes ($5\%$, $10\%$, $20\%$ and $100\%$
of training samples).}
\end{figure*}
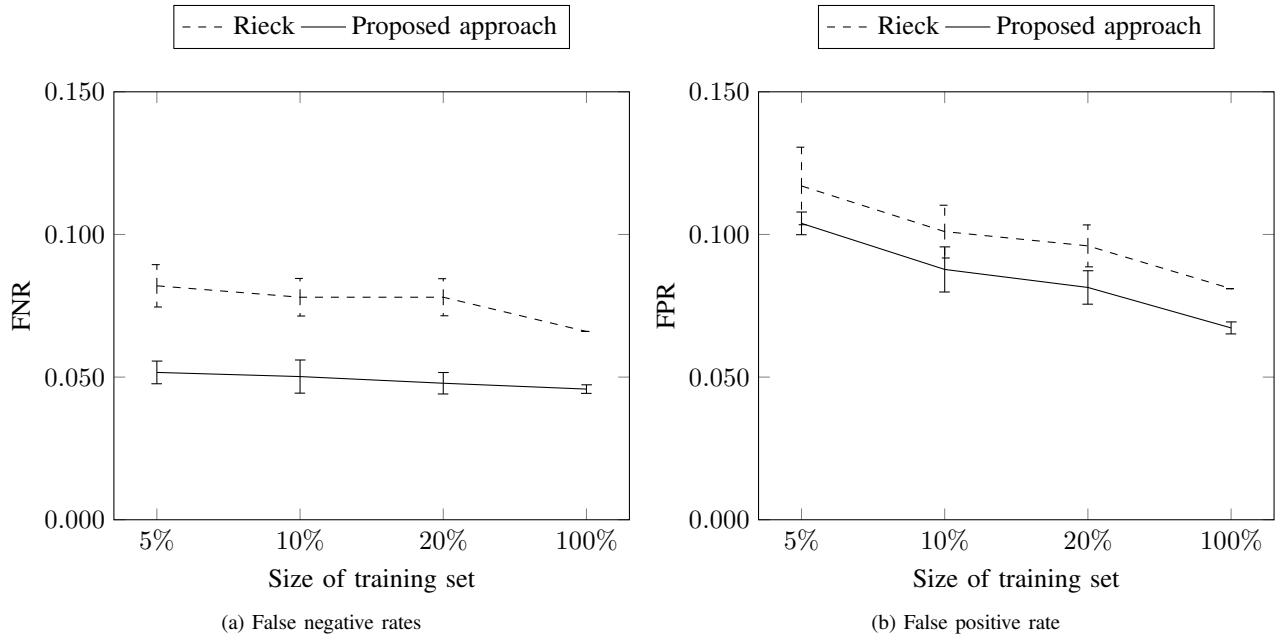
 The detection rates and accuracy of classifiers trained on all $121\,115$
training samples as estimated on testing samples are shown in Table~\ref{tab:results}.
The differences between evaluation metrics indicate that the proposed
approach outperforms both Rieck and AMAL having the lowest false positive
rate and false negative rate. A deeper analysis of the misclassifications
produced by the proposed approach revealed that most of the false
positives (legitimate binaries classified as malware) were software
utilities such as TeamViewer that install themselves into system directories
without any user interaction. Since their incidence in the training
set was relatively low, the random forest was not able to precisely
learn this type of behavior. A second source of errors are false negatives
(malware samples classified as benign) where almost $70\%$ are caused
by insufficient numbers of training samples (less than 100 samples)
from corresponding malware families. Another $11\%$ of false negatives
was caused by concept drift as a portion of testing samples exhibited
different behaviors than training samples, i.e. created files or registry
keys followed different pattern, network communication significantly
different URLs, etc.

Large gaps between training and testing accuracies for AMAL and Rieck
suggest that manually created features and BoW features do not generalize
over longer periods of time as well as features created through clustering
do. This suggests that clustering removes some randomization of resource
names while retaining a large part of information content. 

Figures~\ref{fig:fnr_degradation} and \ref{fig:fpr_degradation}
show graphs of FNR and FPR rates for larger sizes of the training
set expressed as fraction of the data available for training. For
fair comparison the testing set was kept static containing all $110\,140$
samples collected after November 12, 2016. Both graphs show that the
proposed approach is able to achieve lower FNR and FPR using fewer
samples. In fact, the proposed approach achieved FNR of 0.052 using
just 5\% of samples, while Rieck achieved 0.066 using the full training
set. Similarly, the proposed approach needed just 20\% of samples
to achieve the same FPR 0.081 as Rieck on all samples. 

Figures~\ref{fig:fnr_degradation} and \ref{fig:fpr_degradation}
also shows that while false negative rates almost do not change with
respect to the size of the training set (especially for the proposed
approach), the false positive rates decrease dramatically. This suggests
that learning behavior of legitimate applications  is more difficult
than that of malware, which can be caused by the fact that the behavior
of malware is more uniform than that of legitimate applications. This
corroborates the motivation of this work, that even though malware
authors try to randomize, they tend to randomize with same sort of
regularity, which leads to uniformity. 

\subsection{Detection limits}

The experimental results hint at where are the limits of classifying
binaries executed in sandbox. When a binary (or all binaries of some
malware family) does not perform any actions changing the  data used
by the proposed or other methods (files, mutexes, network communication,
registry keys) it clearly evades detection. An example of such malware
is bitcoin miner that resides only in memory without any additional
footprint (no operations with files, no operations with registry keys,
no mutexes, very limited network communication). Such malware has
to be carefully crafted to avoid any interaction with system resources
(statically compiled to carry all libraries in the executable, limited
network communication, no mutexes ensuring that only single instance
is running on the same machine, no persistency after reboot, etc.).
Fortunately, at the time of writing this work, this is not an easy
task and the majority of malware authors choose to interact with system
resources rather than sacrifice functionality.

Another limitation is the fact that a growing number of malware families
are equipped with advanced anti-VM and anti-sandbox features and/or
are targeted to specific environments (Stuxnet~\cite{Falliere2011}).
Such malware families do not reveal their true purpose during sandboxing
or mimic less severe types of malware (adware, PUA\footnote{Potentially unwanted application.},
etc.). This fact is recognized by the community as the main factor
hindering the performance of dynamic analysis as the whole. Addressing
this issue is out of the scope of this paper.

The last aspect we need to discuss is the false positive rate. The
analysis of the results from Section~\ref{sec:Experiments} revealed
that a large number of false alarms is caused by applications that
install themselves into system directories without user's interaction
and since their number is limited, the classifier was unable to fit
this behavior. A solution is of course to improve the training data
by including a larger number of such samples and thus achieve lower
false positive rate. 

\subsection{Scalability and computational complexity}

The last aspect we will discuss is the scalability of the proposed
solution and prior art. Since the proposed solution employs clustering
to project the input data into a feature space with a lower dimension,
a large portion of the training time is spent in the clustering phase.
However, the preprocessing of the dataset used in above experiments
was much faster ($\sim2\text{h}50\text{min}$) than the highly optimized
pre-computation of the full kernel matrix required by Rieck ($\sim7\text{h}$).
This is caused by the fact that the time required by Rieck for preprocessing
grows quadratically with the number of training samples in contrast
to the proposed solution with linear complexity (up to an additive
constant, see Section~\ref{subsec:clusteringDescription}). Moreover,
the proposed solution can be easily distributed since in every iteration
the nearest neighbor search depends only on a limited set of current
cluster prototypes $C'.$

Another benefit of the proposed solution is tied to the representation
itself. Since the clustering is performed only on training samples,
in order to classify unknown samples we need to store only the cluster
prototypes determined during training. For the whole training dataset
used in this paper, which contains over 7 million unique resource
names projected into $\sim40\,000$ features, only $400\,000$ instances
need to be stored. In contrast, the kernelized SVM classifier used
by Rieck et al. requires to store all training samples (over $120\,000$
samples in the data discussed in Section~\ref{sec:Experiments})
with all actions (on average $2000$ actions per sample) in order
to make prediction on unknown samples.

In contrast to both the proposed solution and Rieck, AMAL does not
need any preprocessing since the features can be extracted per sample.
However, the complexity arises from the design of the training process.
Authors in~\cite{Mohaisen2015} argue that the dynamic selection
of both optimal algorithm and its parameters provides optimal results,
but this design makes the training process computationally expensive
since every training of the meta-learner requires to evaluate all
possible combinations of parameters for all its classifiers. Another
aspect is the selection of classifiers itself. Authors propose to
use an array of classifiers such as kernelized SVM, linear regression,
decision trees, perceptron, etc. However, the complexity of some classifiers
(e.g. kernelized SVM) prevents any large-scale training. Moreover,
according to the evaluation the AMAL's detection capabilities are
not sufficient for real-world deployment since both FPR and FNR are
nearly $20\%$, which is clearly insufficient.

\section{Related work}

\label{sec:RelatedWork}Since the analysis of malicious binaries and
recommending them for further analysis has important practical applications,
there exists rich prior art. Although it is frequently divided into
two categories, static and dynamic, the boundaries between them are
blurred since techniques such as analysis of the execution graph is
used in both categories.

\subsection{Static malware analysis}

Static malware analysis treats a malware binary file as a data file
from which it extracts features without executing it. The earliest
approaches~\cite{Lo1995} looked for a manually specified set of
specific instructions \emph{(tell-tale)} used by malware to perform
malicious actions but not used by legitimate binaries. Latter works,
inspired by text analysis, used $n$-gram models of binaries and instructions
within \cite{Li2005}. Malware authors reacted quickly and began to
obfuscate, encrypt, and randomize their binaries, which rendered such
basic models~\cite{Sharif2008} useless. Since reversing obfuscation
and polymorphic techniques is in theory an NP-hard problem~\cite{Moser2007},
most state of the art~\cite{Christodorescu2006,Ahmadi2016,Sharif2008a}
moved to a higher-level modeling of sequences of instructions / system
calls and estimating their action or effect on the operating system.
The rationale behind is that higher-level actions are more difficult
to hide. 

\subsection{Dynamic malware analysis}

An alternative solution to analyzing obfuscation and encryption is
the execution of a binary in a controlled environment and analyzing
its interactions with the operating system and system resources. 

A large portion of the work related to dynamic malware analysis utilize
system calls, since in modern operating systems system calls are the
only way for applications to interact with the hardware and as such
they can reveal malware actions. The simplest methods view a sequence
of system calls as a sequence of strings and use histograms of occurrences
to create feature vectors for the classifier of choice~\cite{Hansen2016}.
The biggest drawback of these naive techniques is low robustness to
system call randomization. Similarly to static analysis, this problem
can be tackled by assigning actions to groups (clusters) of system
calls (syscalls) and using them to characterize the binary~\cite{Naval2015,Wuchner2014,Bayer2009}. 

A wide class of methods identifying malware binaries from sequences
of syscalls rely on $n$-grams~\cite{Lanzi2010,OKane2013}. Malheur~\cite{Rieck2011}
uses normalized histograms of $n$-grams as feature vectors, which
effectively embeds syscall sequences into Euclidean space endowed
with $L_{2}$ norm. In this space\textcolor{black}{{} the algorithm
extracts prototypes $Z=\left\{ z_{1},\dots,z_{n}\right\} $ using
hierarchical clustering. Each prototype captures the behavior of the
cluster, which should match corresponding malware family. An interesting
feature of Malheur is that if a cluster has less then a certain number
of samples, the prototype is not created. The classification of an
unknown binary is determined by searching for the nearest prototype
within certain range. If the nearest prototype is outside of this
range, the sample is not classified.}

To counter dynamic analysis advanced malware detects the presence
of a sandbox and does not execute within it. Since most sandboxes
rely on a detectable system call interposition,  Das et al.~\cite{Das2016}
propose to extend hardware with FPGA that would extract system calls
from their execution on processor. Syscalls are then grouped by comprehensive
yet hand designed rules, and these groups are then fed into multi-layer
neural network classifier. The classifier itself is also part of the
FPGA, such that the system can simultaneously extract training samples
and classify them.

AMAL uses its custom sandbox to extract features describing files,
network communication and registry features~\cite{Mohaisen2015}
and tunes various classification algorithms. The main difference between
AMAL and this work is the construction of features. Whereas AMAL uses
numeric features such as counts or sizes of created, modified or deleted
files, counts of created, modified or deleted registry keys, counts
of unique IP addresses, etc., we assume that individual resources
(files, registry keys, mutexes and network communication) have specific
role in the operation system, which can be different even though the
characteristics exhibited by the file are the same.

The approach proposed by Rieck et al. \cite{Rieck2008} creates a
representation of the analyzed binaries directly from the data which
is at the first sight similar to the proposed approach, however there
are two key differences. The first one is the source of data, because
Rieck et al. model actions triggered by the malware (writing into
a file, communication with remote server, reading data from registry
keys, starting new thread, etc.), whereas the proposed approach models
only affected resources. This enables to deploy the proposed approach
in environments without access to low-level actions (VMs without such
access, user machines without API hooking). Another difference is
in handling the randomization of resource names. Instead of clustering
resource names used in this work, Rieck et al. remove parameters of
actions, which increases the dimensionality of the model since for
every action with $n$ parameters it creates $n+1$ features representing
the action at different levels of granularity by removing parameters
from the end: from full description with all parameters to the most
coarse description where only name of the action is used. This leads
to a massive increase in the already large number of features.\footnote{According to the experiments, the number of features generated for
about 6000 samples reaches over 20 million.} Even though the resulting feature space, is sparse the scalability
of such an approach is limited.

\section{Conclusion}

\label{sec:Conclusion}Dynamic malware analysis is a popular approach
to automatically identify malware binaries and analyze them. This
paper has proposed a model of malware behavior observed through its
interactions with the operating system and network resources (operations
with files, mutexes, registry keys, operations with network servers
or error messages provided by the operating system). It employs an
efficient clustering of resource names to reduce the impact of randomization
commonly employed by malware authors to avoid detection and projects
malware samples into a low-dimensional space suitable for classifiers
such as random forest.

The proposed solution was extensively compared to related state of
the art on a large corpus of binaries where it demonstrated significant
increase in precision of malware detection. Moreover, we believe that
the availability of solutions relying on widely different types of
features increases the overall reliability of malware detection techniques,
because malware authors have to evade more detectors to stay undetected.

\bibliographystyle{plain}
\bibliography{library}

\appendix

\section{Appendix}

\subsection{Frobenius product and Frobenius norm\label{subsec:Frobenius-product}}

For two matrices $\mathbf{A}\in\mathbb{R}^{n\times m}$ and $\mathbf{B}\in\mathbb{R}^{n\times m}$
we define \emph{Frobenius product $\left\langle \cdot,\cdot\right\rangle _{F}$
}and \emph{Frobenius norm }$\left\Vert \cdot\right\Vert _{F}$ as
follows

\begin{align*}
\left\langle A,B\right\rangle _{F}=\sum_{i=1}^{n}\sum_{j=1}^{m}A_{ij}\cdot B_{ij}
\end{align*}

\begin{align*}
\left\Vert \mathbf{A}\right\Vert _{F}=\sqrt{\left\langle A,A\right\rangle _{F}}=\sqrt{\sum_{i=1}^{n}\sum_{j=1}^{m}A_{ij}^{2}}
\end{align*}

\end{document}